\documentclass{elsart}
\usepackage{natbib}
\usepackage{graphicx}
\def\lsim{\lower.5ex\hbox{$\; \buildrel < \over \sim \;$}}
\def\gsim{\lower.5ex\hbox{$\; \buildrel > \over \sim \;$}}
\journal{New Astronomy Reviews}
\begin{document}
\runtitle{Gamma-ray lines from interstellar dust grains}
\begin{frontmatter}
\title{Gamma-ray lines from cosmic-ray interactions with interstellar
dust grains}
\author{Vincent Tatischeff and J\"urgen Kiener}
\address{Centre de Spectrom\'etrie Nucl\'eaire et de Spectrom\'etrie 
de Masse, IN2P3-CNRS and Universit\'e Paris-Sud, F-91405 Orsay Cedex, France}
\begin{abstract}
As pointed out by \citet{L77}, the shapes of some
$\gamma$-ray lines produced by cosmic-ray interactions with the
interstellar medium potentially contain valuable information on the
physical properties of dust grains, including their compositions and
size distributions. The most promising of such lines are at 847,
1369, 1779 and 6129~keV, from $^{56}$Fe*, $^{24}$Mg*, $^{28}$Si* and
$^{16}$O*, respectively. We performed detailed calculations of their
profiles using, in particular, available laboratory measurements combined 
with optical model calculations to evaluate the energy distributions of 
the recoiling excited nuclei. We show that the line shapes are mainly 
sensitive to relatively large interstellar grains, with radii 
$\gsim$0.25~$\mu$m. Line fluxes from the inner Galaxy are then predicted. 
\end{abstract}
\end{frontmatter}

\section{Introduction}

Observations of nuclear interaction $\gamma$-ray lines
from the interstellar medium (ISM) would provide a unique tool to study
Galactic cosmic-ray ions at non-relativistic energies, as well as the physical 
conditions of the emitting regions. If the lines produced in the gaseous 
phase are expected to be significantly Doppler-broadened, 
some lines produced in interstellar dust grains can be very narrow, because 
some of the excited nuclei can stop in solid materials before emitting 
$\gamma$-rays \citep{L77,R79}. The latter are prime candidates for detection 
with $\gamma$-ray telescopes having high spectral resolution, such as the 
{\em INTEGRAL} spectrometer (SPI). 

\begin{figure*}
\begin{center}
\includegraphics[width=13.1cm]{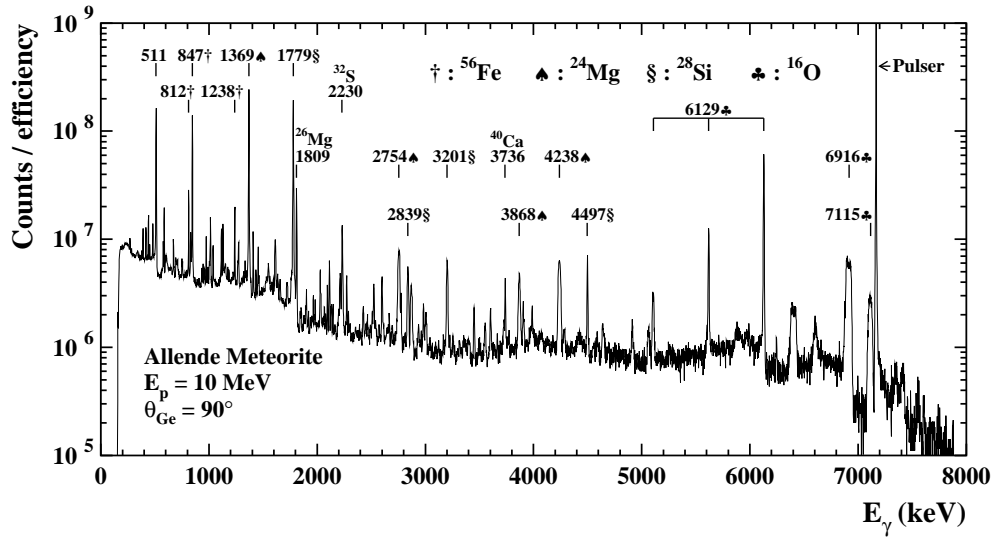}
\end{center}
\caption{Observed $\gamma$-ray spectrum from the bombardment of a thick
sample of the Allende meteorite with 10-MeV protons. The most intense lines 
are labeled with their nominal energies and the target nuclei from which the 
$\gamma$-rays are produced. For the 6129-keV line, the single and double 
escape peaks are also indicated.}
\label{f1}
\end{figure*}

An illustrative $\gamma$-ray spectrum is shown in figure (\ref{f1}).
It was obtained in a recent experiment performed at Orsay (Kiener et
al., in preparation), in which a thick sample of the Allende meteorite 
($\sim$5~mm in diameter) was bombarded with 10-MeV protons. The Allende 
meteorite belongs to the class of carbonaceous chondrites and has a 
composition close to cosmic, except for the volatile elements. The strongest 
lines (apart from the 511-keV emission) originate from reactions with the 
most abundant isotopes: $^{16}$O, $^{24}$Mg, $^{28}$Si and $^{56}$Fe. One 
point of particular interest is that the $^{16}$O* line at 6129~keV 
is much narrower than the two other $^{16}$O* lines at 6916 and 7115~keV. 
This is because the 6.13-MeV state is relatively long-lived 
($T_{1/2}$=18.4~ps), such that the recoiling $^{16}$O* nuclei can come to
rest in the target before the 6.129-keV $\gamma$-ray is emitted, whereas 
the 6.92- and 7.12-MeV states ($T_{1/2}$=4.7 and 8.3~fs, respectively) mostly 
de-excite in flight. The 6129-keV line, as well as the intense lines at 
847~keV from $^{56}$Fe* ($T_{1/2}$=6.1~ps), 1369~keV from $^{24}$Mg* 
($T_{1/2}$=1.35~ps) and 1779~keV from $^{28}$Si* ($T_{1/2}$=475~fs) are the 
prime candidates to scrutinize the interstellar dust grains through their 
$\gamma$-ray emission. 

We have calculated in detail the shapes of these four lines as they are 
produced by cosmic-ray interactions with the ISM. The interaction model and 
the results are presented in sections (\ref{s2}) and (\ref{s3}), 
respectively. In section (\ref{s4}), we evaluate the intensities of the 
predicted emissions from the inner Galaxy. 

\section{Interaction model}
\label{s2}

\begin{figure*}
\begin{center}
\includegraphics[width=10.5cm]{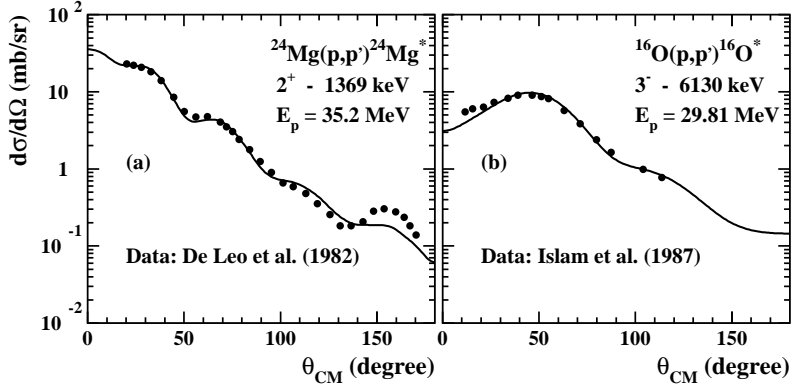}
\end{center}
\caption{Differential inelastic scattering cross sections for the
reactions (a) $^{24}$Mg($p$,$p'$)$^{24}$Mg$^*_{1369}$ at
$E_p$=35.2 MeV and (b) $^{16}$O($p$,$p'$)$^{16}$O$^*_{6130}$ at
$E_p$=29.81 MeV. Solid curves -- coupled-channel calculations.}
\label{f2}
\end{figure*}

To calculate the line profiles, we used a Monte-Carlo method
similar to that described in \citet{R79}. Each event in the simulation
corresponds to a $\gamma$-ray producing nuclear reaction of a cosmic-ray 
proton or $\alpha$-particle with a target nucleus either in the gas or
locked up in a spherical and homogeneous dust grain of the ISM. 
The $\gamma$-rays can be produced by inelastic scattering reactions, e.g.
$^{16}$O($p$,$p'$$\gamma_{6129}$)$^{16}$O, by $(p,n)$ reactions
followed by electron capture or positron emission, e.g.
$^{56}$Fe$(p,n)$$^{56}$Co($\epsilon$-$\beta^+$)$^{56}$Fe$^*_{847}$
($T_{1/2}$($^{56}$Co)=77.2~d), or by spallation reactions,
e.g. $^{20}$Ne($p$,$p$$\alpha$$\gamma_{6129}$)$^{16}$O. The total cross 
sections are from \citet{K02}, except for the reactions 
$^{24}$Mg$(p,n)$$^{24}$Al($\beta^+$)$^{24}$Mg$^*_{1369}$
($T_{1/2}$($^{24}$Al)=2.053~s) and 
$^{28}$Si$(p,n)$$^{28}$P($\beta^+$)$^{28}$Si$^*_{1779}$
($T_{1/2}$($^{28}$P)=270.3~ms). We estimated the contribution of the
$^{24}$Mg$(p,n)$$^{24}$Al channel from the data of \citet{K89} and assumed 
the same relative contribution for the reaction $^{28}$Si$(p,n)$$^{28}$P. 

The energy distribution of the recoiling excited nuclei were calculated from 
the differential cross sections. The latter were obtained 
from a large number of experimental data, completed with
extensive coupled-channel calculations with the code ECIS94 \citep{R94}. 
Two examples of differential inelastic scattering cross sections are shown 
in figure (\ref{f2}). 

The stopping powers of the nuclei recoiling in the 
grain material were calculated with the code TRIM \citep{ZIE}. For
these calculations, the grain composition was assumed to be
(MgSiFe)O$_4$, which is characteristic of interstellar silicates
\citep[e.g.][]{D97}. 

\begin{figure}
\begin{center}
\includegraphics[width=6.7cm]{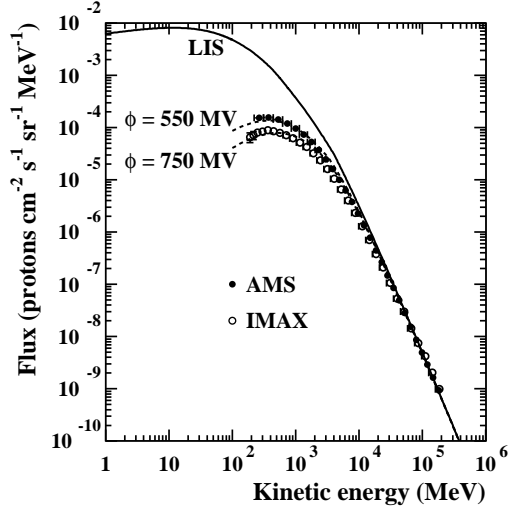}
\end{center}
\caption{Cosmic-ray proton local interstellar spectrum (LIS;
solid curve) and solar-modulated spectra for two force-field potentials, 
$\phi$=550 and 750 MV (dashed curves). AMS: \citet{AMS}; IMAX: \citet{IMAX}.}
\label{f3}
\end{figure}

To estimate the cosmic-ray proton interstellar spectrum, we used the
disk-halo propagation model of \citet{J01}. Coulomb and ionization
energy losses were taken from \citet{M94}. For the proton inelastic
cross sections, we used the empirical formula given in \citet{M02}. 
Solar-modulated spectra, calculated from the force-field 
approximation, were fitted to measured proton fluxes (fig. \ref{f3}). We 
found the best fit to be provided by the source spectrum 
$\dot{Q}(p)dp \propto p^{-2.4}dp$, where $p$ is the proton momentum. For
simplicity, we used the same form for the $\alpha$-particle LIS spectrum as 
for the protons, with an abundance ratio $\alpha/p$=0.1. 

\section{Gamma-ray line profiles}
\label{s3}

\begin{figure*}
\begin{center}
\includegraphics[width=11.5cm]{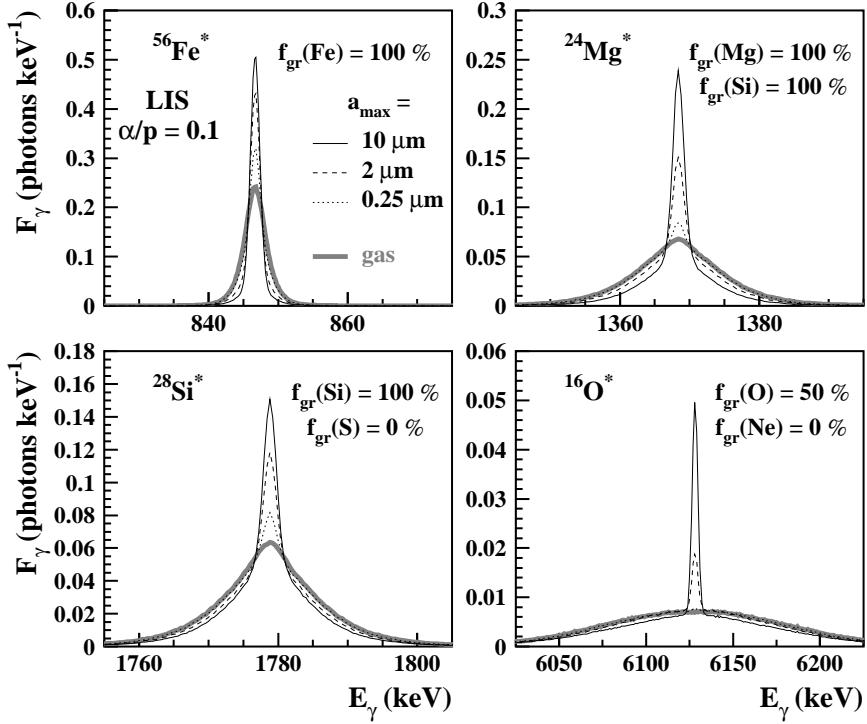}
\end{center}
\caption{Profiles of the $\gamma$-ray lines at 847~keV ($^{56}$Fe*),
1369~keV ($^{24}$Mg*), 1779~keV ($^{28}$Si*) and 6129~keV ($^{16}$O*), 
excited in cosmic-ray proton and $\alpha$-particle interactions with 
interstellar gas and dust grains. The calculated spectra are normalized to
one photon emitted in each line and convolved with the SPI response function. 
Thin curves -- grain size distributions following an $a^{-3.5}$ power 
law in radii from $a_{min}$=5~nm to $a_{max}$=0.25~$\mu$m (dotted curves), 
2~$\mu$m (dashed curves) and 10~$\mu$m (solid curves). For each $\gamma$-ray
line, the fractions of target nuclei assumed to be in the grains are 
indicated. Thick, grey curves -- all target nuclei are assumed to be in the 
gas.}
\label{f4}
\end{figure*}

Calculated $\gamma$-ray line profiles are shown in figure (\ref{f4})
for different interstellar grain size distributions. We
assumed that all the available refractory elements Mg, Si and Fe, and half
of the O are locked up in silicate grains, whereas the volatile elements Ne
and S (which contribute through spallation reactions to the 6129- and 1779-keV
lines, respectively) are in the gaseous phase \citep[see][]{S96,D97}.
The dotted curves were obtained for the commonly used MRN \citep{M77}
size distribution: an $a^{-3.5}$ power law in grain radii $a$ from 
$a_{min}$=5~nm \citep[see][]{D97} to $a_{max}$=0.25~$\mu$m. We see 
that there are little differences between these spectra and those calculated 
assuming that all the target nuclei are in the interstellar gas (thick, grey 
curves). In particular, the two profiles of the 6129-keV line are 
nearly identical, because almost all of the $^{16}$O nuclei excited within 
dust grains of radius $<$0.25~$\mu$m escape from the solids before decaying. 

The dashed curves were calculated for the same power law distribution, 
but with $a_{max}$=2~$\mu$m. The extension of the grain radii beyond the MRN
limit of 0.25~$\mu$m is motivated by the observations with dust detectors
aboard the {\em Ulysses} and {\em Galileo} spacecrafts of relatively large
interstellar grains entering the solar system \citep{L00}, as well as by
models of scattering halos observed around X-ray sources \citep{W01}. The
thin, solid curves were obtained by further extending the MRN distribution 
to $a_{max}$=10~$\mu$m. Such large grains are typical of presolar grains of
stardust found in primitive meteorites \citep{Z98} and should exist in
various circumstellar media. We see that for these two grain size 
distributions, the very narrow component of the $^{24}$Mg*, $^{28}$Si* and
$^{16}$O* lines could in principle be resolved with SPI. The sensitivity of
these line shapes to micrometer-sized particles provides a promising method
for tracing large dust grains in the ISM. 

\section{Gamma-ray line fluxes from the inner Galaxy}
\label{s4}

The $\gamma$-ray line intensities from the inner Galaxy were estimated by
normalizing the emissivity calculations to the observed flux of high-energy
$\gamma$-rays ($>$70~MeV) due to $\pi^0$ production and decay. We derived
the latter from \citet[][fig.~1; EGRET data of observing phases 1-4]{B01}:
$\Phi_\gamma^{\pi^0}$$\simeq$8.3$\times$10$^{-5}$ photons cm$^{-2}$ s$^{-1}$, 
for -40$^\circ$$<$$\ell$$<$40$^\circ$ and -6$^\circ$$<$$b$$<$6$^\circ$. We 
used the model of \citet{D86} to calculate the production of $\pi^0$-decay
$\gamma$-rays by cosmic-rays with the LIS spectrum of figure (\ref{f3}). We 
assumed that the metal abundances in the inner Galaxy are on the average twice 
solar. We then obtained for the fluxes of the 847-, 1369-, 1779- and 6129-keV 
lines excited in proton and $\alpha$-particle interactions with both 
interstellar gas and dust grains: 5.6$\times$10$^{-8}$, 7.5$\times$10$^{-8}$,
4.3$\times$10$^{-8}$ and 3.1$\times$10$^{-7}$ photons cm$^{-2}$ s$^{-1}$,
respectively. In comparison, the calculated flux of the relatively strong
line at 4438 keV from $^{12}$C* (FWHM$\simeq$150~keV) is 7.1$\times$10$^{-7}$
photons cm$^{-2}$ s$^{-1}$ and that of the so-called $\alpha$$-$$\alpha$ line
at $\sim$450~keV (complex line shape with FWHM$\simeq$80~keV; see
Tatischeff et al., 2001) is 3.9$\times$10$^{-7}$ photons cm$^{-2}$ s$^{-1}$. 
All these diffuse emission fluxes are far below the SPI sensitivity, such 
that a near future detection of nuclear interaction $\gamma$-ray lines from 
the inner Galaxy is unlikely, unless there is a large population of Galactic 
cosmic-rays with kinetic energies below the threshold for $\pi^0$ production 
(290~MeV for $p$+H collisions). Such a distinct cosmic-ray component 
predominant at low energies has been proposed to account for the quasi 
linear correlation between Be and Fe abundances for metallicities 
[Fe/H]$<$-1 \citep[e.g.][]{C95}.

\ack{We acknowledge F. Boulanger for usefull discussions.}

\end{document}